\documentclass[prd,amsmath,amssymb,superscriptaddress,twocolumn,nofootinbib,floatfix,showpacs]{revtex4}
\input{epsf}
\usepackage{epsf}
\usepackage{graphicx,epsfig}
\usepackage{bm}
\usepackage{latexsym,float}

\newcommand {\ga} {\ {\raise-.5ex\hbox{$\buildrel>\over\sim$}}\ }
\newcommand {\la} {\ {\raise-.5ex\hbox{$\buildrel<\over\sim$}}\ } 

\newcommand{\fig}[1] {Fig.~(\ref{#1})}

\def\be{\begin{equation}}
\def\ee{\end{equation}}
\def\ba{\begin{eqnarray}}
\def\ea{\end{eqnarray}}
\renewcommand{\(}{\left(} 
\renewcommand{\)}{\right)} 
\renewcommand{\[}{\left[} 
\renewcommand{\]}{\right]}

\renewcommand{\Psi}{\varPsi}

\begin{document}

\title{Dark radiation as a signature of dark energy}  

\author{Sourish~Dutta}
\email{sourish.d@gmail.com}
\affiliation{Department of Physics and Astronomy, Vanderbilt University,
Nashville, TN 37235}

\author{Stephen~D.~H.~Hsu} \email{hsu@uoregon.edu}
\affiliation{Institute of Theoretical Science, University of Oregon,
Eugene, OR 97403-5203}

\author{David~Reeb} \email{dreeb@uoregon.edu}
\affiliation{Institute of Theoretical Science, University of Oregon,
Eugene, OR 97403-5203}

\author{Robert~J.~Scherrer}
\email{robert.scherrer@vanderbilt.edu}
\affiliation{Department of Physics and Astronomy, Vanderbilt University,
Nashville, TN 37235}

\begin{abstract}
We propose a simple dark energy model with the following properties:
the model predicts a late-time dark radiation component that is
not ruled out by current observational data, but which produces
a distinctive time-dependent equation of state $w(z)$ for $z < 3$.
The dark energy field can be coupled strongly enough to
Standard Model particles to be detected in colliders, and
the model requires only modest additional particle content
and little or no fine-tuning other than a new energy scale
of order milli-electron volts.
\end{abstract}


\date{May 6, 2009}

\maketitle

\section{Introduction} 
\label{Introduction}

Considerable evidence \cite{Knop,Riess1} has accumulated 
suggesting that approximately 70\% of the energy density in the
Universe comes in the form of an exotic, negative-pressure component,
called dark energy. (For a recent
review, see \cite{Copeland}.)

The equation of state (EoS) parameter is defined as the ratio of the dark energy pressure to its density:
\be
w=p_{\rm DE}/\rho_{\rm DE}~.
\ee
Observations constrain $w$ to be very close to $-1$. For instance, if $w$ is assumed to be constant, then necessarily $-1.1 \la w \la -0.9$  \cite{Wood-Vasey,Davis}. If $w=-1$, the dark energy density remains constant even though the Universe is expanding. The simplest way of producing a $w=-1$ component is through a cosmological constant, or vacuum energy density. However, as is well known, the energy density needed to explain the observed acceleration, $\Delta^4\equiv\(10^{-3}\,\text{eV}\)^4$, is considerably smaller than the value of $\(10^{19}\,\text{GeV}\)^4$ (Planck density) predicted from quantum field theory. This $124$-orders-of-magnitude discrepancy is called the cosmological constant problem. 

The fact that the observed vacuum energy also happens to be just a few times greater than the present matter density has led to speculations that it might in fact be evolving with time -- only now reaching a value comparable to the matter density. Such a time-varying vacuum energy is sometimes referred to as \emph{quintessence}. The simplest way of achieving a time-varying vacuum energy is through the use of spatially homogeneous canonical scalar fields \cite{RatraPeebles,TurnerWhite,CaldwellDaveSteinhardt,LiddleScherrer,SteinhardtWangZlatev}. In these models, the field typically rolls down a very shallow potential, eventually coming to rest when it can find a local minimum.

Quintessence models typically have fine-tuning problems. For example, since the quintessence redshifts more slowly than ordinary matter or radiation, the current quintessence dominance can only be explained by fine-tuning the initial conditions. This problem can be avoided in the class of so-called ``tracker'' models, in which the evolution of the quintessence field is insensitive to the initial conditions. For generic quintessence models,  the flatness of the potential makes any excitations of the field almost massless $\sim10^{-33}\,\text{eV}$. To provide the necessary vacuum energy density, the present value of the potential energy should be on the order of $\Delta^4$ (although there is really no rigorous physical reason to expect this). The field value $\phi_0$ today should therefore be on the order of the Planck mass, i.e.~$\phi_0\sim 10^{18}\,\text{GeV}$.\footnote{An alternative class of models which relies on non-linear oscillations of the quintessence field and does not require an extremely flat potential was proposed in \cite{Sahni,Hsu} and discussed further in \cite{Masso,Gu,DuttaScherrer,Johnson}.} In \cite{Carroll} it was shown that couplings between quintessence and ordinary matter, even if Planck-suppressed, can lead to long range forces and time-dependence in the constants of Nature, both of which are tightly constrained. Reference \cite{HsuMurray} showed that even Planck-suppressed thermal interactions between matter and quintessence can significantly alter the evolution of the latter, leading to a problematic equation of state. 

An alternative to a slowly rolling field is a scenario where the field is stuck in a false vacuum minimum. In this case, the observed cosmological constant is attributed to the energy difference between the false and true vacua, which could either arise from higher-order \cite{Garretson:1993kg,Barr:2001vh} or non-perturbative effects \cite{Yokoyama:2001ez}. Other proposals for the origin of the false vacuum energy are the confining scale of a hidden $SU(2)$ sector \cite{Goldberg:2000ap}, Planck-scale suppressed mediation into a hidden sector of electroweak TeV-scale supersymmetry breaking \cite{ArkaniHamed:2000tc,Chacko:2004ky}, or the vacuum energy of a hidden sector which is stuck in a state of equilibrium between phases \cite{Megevand}. 

In many of these quintessence models, the field(s) responsible for the acceleration have to be almost completely decoupled from the rest of the Universe.\footnote{For an entirely different type of dark energy model, which can have a particle physics signature, see \cite{Nelson}. Here the acceleration is provided by mass-varying neutrinos (MaVaNs) which act as a negative-pressure fluid.} This is disappointing, since it suggests that direct detection of quintessence through its interactions with Standard Model particles will be extremely challenging, perhaps impossible. 

In this paper we present a quintessence scenario in which the dark energy field can be coupled strongly enough to Standard Model particles to be detected in colliders, and which allows for a significant time variation in the equation of state. This time-varying $w=w(z)$ has a characteristic form which depends on only a single parameter, and can thus be excluded by cosmological observations in the near future. Our model only requires a singlet scalar field (or, alternatively, a small gauge sector like $SU(3)$ Yang-Mills theory; other possible realizations are also briefly outlined at the end of the paper) and a new energy scale on the order of milli-electron volts. 

\section{Model}
\label{aa}
\begin{figure}
\epsfig{file=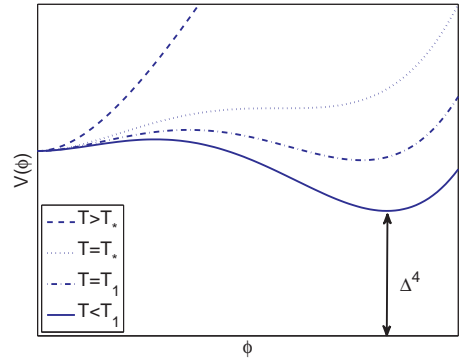,height=52mm}
\caption{\label{potential}An example for the evolution of the finite temperature effective potential $V\(\phi,T\)$ of the dark energy field, eqn.~(\ref{effpot}), as the temperature $T$ decreases through the first order phase transition region $\sim T_1$.}
\end{figure}

Consider a singlet scalar field dark energy with Lagrangian
\begin{equation}
\label{L}
{\cal L} = \frac{1}{2}\(\partial_\mu \phi\)^2 - V \(\phi\)~.
\end{equation}
We allow this field to be strongly coupled to Standard Model particles. The finite temperature effective potential, which includes interactions of this field with virtual particles and the heat bath, can be taken to be similar to the Higgs potential in the electroweak phase transition (see, e.g., \cite{AndersonHall} for a review):
\begin{equation}
V \(\phi, T\) =A+D\(T^2-T_{2}^2\)\phi^2  - E T \phi^3 + \frac{1}{4}\lambda \phi^4~.\label{effpot}
\end{equation}
$D$, $E$, $\lambda$ and $A$ are constants. $A$ can be adjusted to give the correct value of the observed dark energy density when $T=0$.
$T_2$ is defined as the temperature where $V''\(\phi=0\)=0$. We choose $T_2=\Delta$, i.e.~roughly $T_2~\sim11.6\,\text{K}$, and assume that it represents a new energy scale in particle physics. At high temperatures, $T\gg T_2$, $\phi=0$ is the only minimum of the potential. As the Universe cools down, an inflection point appears in the potential at temperature $T_*=T_2/\sqrt{1-9 E^2/8\lambda D}$. At lower temperatures, this splits into a barrier and a second minimum. The critical temperature $T_1=T_2/\sqrt{1- E^2/\lambda D}$  corresponds to the point where the second minimum is equal in (free) energy to the $\phi=0$ minimum. At temperatures $T<T_1$, the second minimum has lower free energy than the one at $\phi=0$. The evolution of the potential well with temperature is shown in \fig{potential}.

Let us now consider the dynamics of the dark energy field. At temperatures $T> T_1$ (which corresponds to roughly $z>3$) the dark energy field remains trapped at the $\phi=0$ minimum, providing a constant energy density, which we assume to be slightly higher than $\Delta^4$. As the temperature approaches $T_1$ and below, a first order phase transition is triggered as the field tunnels into the true vacuum $\Delta$. The physics of the phase transition is almost identical to that of the Higgs sector in models of electroweak baryogenesis.  This transition releases energy in relativistic modes (i.e., scalar particles of the $\phi$ field), and brings the vacuum energy to $\Delta$. Because the correlation length of the transition is microscopic, and the relativistic modes couple weakly to ordinary matter (i.e., more weakly than photons, perhaps similar to neutrinos), such a transition is only loosely constrained by observation. The positive pressure of the radiation, which eventually redshifts away, causes the effective EoS of the dark energy to vary in (redshift) time $z$.

We note that the only important feature of the model described above is that it has a weakly first order phase transition at a temperature of order $\Delta$, which is natural if one assumes the dynamics of $\phi$ to be entirely determined by that energy scale and dimensionless couplings of order one. It is an interesting coincidence that this occurs at a redshift of $z \sim 3$ if the temperature of the dark energy field is similar to that of the Standard Model particles. This need not be the case, but it seems a reasonable assumption, especially if there are non-negligible interactions between $\phi$ and ordinary particles, which would enforce thermal equilibrium at sufficiently high temperatures. When the transition happens at $z \sim 3$ the resulting radiation component leads to significant and characteristic variation in $w(z)$. The form of $w(z)$ is determined by a single parameter -- the energy fraction in relativistic dark radiation modes just after the phase transition. In some cases, such as the gauge models discussed below, even this fraction is calculable from the phase diagram.

Any sector which produces a weakly first order transition at a temperature of order $\Delta$  would also suffice. For example, pure $SU(N)$ gauge theories with $N > 2$ have first order deconfinement phase transitions \cite{gauge} and exhibit effective potentials like those in Fig.~(\ref{potential}), with $\phi$ an order parameter for confinement, for example the Polyakov loop. Here, the latent heat and fraction of energy in relativistic modes is calculable via lattice simulation.

We stress that the models discussed do not in any way {\it explain} the existence of the energy scale $\Delta$, or why it determines the vacuum energy density today. In particular, why should the vacuum energies from all the other degrees of freedom cancel out, leaving the dark energy field to determine the cosmological constant? One way of explaining this would be to assume that somewhere in the configuration space, outside the region depicted in \fig{potential}, the potential reaches a global minimum $V\(\phi_* \)=0$, where the total vacuum energy (including zero point energies and radiative corrections from {\it all} fields) is exactly zero. That is, some currently unknown mechanism (Euclidean wormholes, quantum gravity, ...) conspires to make the total vacuum energy zero at $\phi = \phi_*$, implying that the deviation of $V \( \phi \)$ from zero is the only vacuum energy.

In any case, if one assumes that new physics at the energy scale $\Delta$ determines the observed cosmological constant, it is easy to obtain a predictable redshift-dependent $w=w(z)$ together with interesting particle physics signatures -- no fine tuning of parameters is required. 

\section{Observational Consequences}\label{Observational Consequences}\subsection{Astrophysics}\vspace{-1mm}As discussed in the previous section, our model produces a certain amount of dark radiation at redshift $z\sim 3$. This radiation affects the Hubble expansion rate $H$ as well as the effective equation of state of the dark energy. 

Let $f$ denote the fraction of the dark energy that is today in the form of relativistic modes, and let us assume that the phase transition occurred at redshift $z_{\rm PT}=3$. The Hubble expansion rate $H\(z\)$ after the phase transition $z<z_{\rm PT}$ can be written as 

\begin{equation}
\begin{split}
\label{hubble1}
H^2\(z\)= H_0^2\[\Omega_{m0}\(1+z\)^3+\Omega_{r0}\(1+z\)^4+\right.\\
\left.+f\Omega_{\phi 0}\(1+z\)^4+\Omega_{\phi 0}\]~,
\end{split}
\end{equation}
where $\Omega_{m0}$, $\Omega_{r0}$ and $\Omega_{\phi 0}$ denote the present-day values of the density parameters of matter, radiation and dark energy. Note that at sufficiently low temperatures the non-zero mass of the dark radiation (coming, e.g., from the curvature at the lower minimum in \fig{potential}) will be non-negligible and its energy density will then redshift as $(1+z)^3$.

\begin{figure}
\epsfig{file=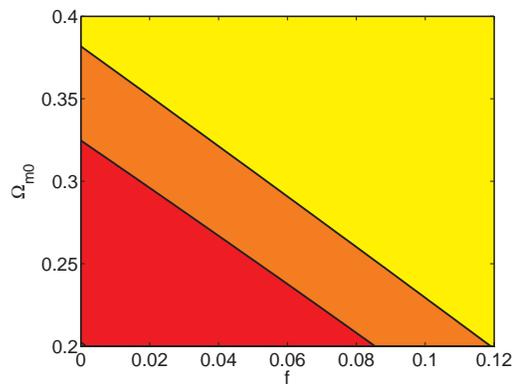,height=52mm}
\caption{\label{likelihood}Likelihood contour for the parameters $f$ and $\Omega_{m0}$. The yellow (light) region
is excluded at the 2$\sigma$ level and the orange (darker) region
is excluded at the 1$\sigma$ level.  The red (darkest) region is
not excluded at either confidence level.}
\end{figure}

Our model superficially resembles other models with a dark radiation component,
such as models with extra relativistic degrees of freedom or the
Randall-Sundrum model with dark radiation.  The difference, of
course, is that in our model the dark radiation arises very late, and so is not
subject to the well-known limits from Big Bang nucleosynthesis or the cosmic
microwave background.  It was noted by Zentner and Walker \cite{ZW} that if one
considers only late-time constraints on extra relativistic degrees of freedom
from SNIa data, the limits are surprisingly weak.  Our results, which we describe now, agree with this
conclusion, even with the addition of more recent SNIa data.

In \fig{likelihood} we construct a likelihood plot for the parameters $\Omega_{m0}$ and $f$. We choose $\Omega_{\phi0}=0.7$ and marginalize over the present value of the Hubble parameter $H_{0}$ using the recent Type Ia Supernovae standard candle data (ESSENCE+SNLS+HST) from \cite{Davis}. Clearly, the SNIa data do not rule out a sizable fraction of the dark energy today being in relativistic modes.

\begin{figure}
\epsfig{file=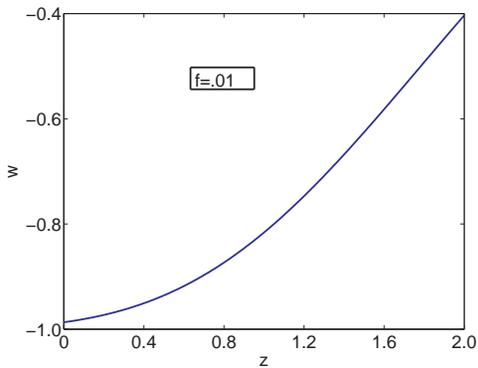,height=52mm}
\caption{\label{wz} $w$ vs.~$z$ for the choice $f=0.01$, which along with, e.g., a conservative $\Omega_{m0}=0.28$ is not excluded at the $1\sigma$ level by the SNIa data (see \fig{likelihood}).}
\end{figure}

It is easy to derive an analytic expression for $w(z)$ in this model.
Taking $p_{DR}$, $\rho_{DR}$ to be the dark radiation pressure and density,
and $p_{\phi}$, $\rho_{\phi}$ to be the scalar field pressure and density, we have
$w = (p_{DR} + p_{\phi})/(\rho_{DR} + \rho_{\phi})$.  But $p_{DR} = \rho_{DR}/3$
and $p_\phi = - \rho_{\phi}$, leading to
\begin{equation}
\label{w(z)}
w(z) = \frac{(1/3)f(1+z)^4 - 1}{f(1+z)^4 + 1}~.
\end{equation}
As an example of the possible strong late-time variation of $w(z)$ predicted by this model, in \fig{wz} we plot $w$ vs.~$z$ for $f=0.01$; e.g., the parameter choice $f=0.01$, $\Omega_{m0}=0.28$ is conservative, it is not excluded at $1\sigma$ by the SNIa data. The pressure of the relativistic component increases the effective EoS of the dark energy component with increasing redshift. Note that, in our model, the shape of the $w$ vs.~$z$ curve is fixed once $f$ is fixed. 

Another diagnostic for dark energy models is the evolution in the $w - w^\prime$ phase plane \cite{CL}, where $w' \equiv a (dw/da)$ and $a  = 1/(1+z)$ is the scale factor. From (\ref{w(z)}), we find that
\begin{equation}
\label{wprime}
w^\prime = (1+w)(3w-1)~.
\end{equation}
Note that the relationship between $w^\prime$ and $w$ is independent of $f$.
This means that all of these models evolve along the same evolutionary
track in the $w - w^\prime$ plane; the value of $f$ simply determines where
the model sits on this evolutionary path at the present time.

In the terminology of \cite{CL}, these are ``freezing'' quintessence models,
since $w$ decreases with time to $w\to-1$.  However, eqn.~(\ref{wprime}) predicts behavior that is distinct from standard freezing quintessence models.  For them, \cite{CL} suggested the bound $w^\prime > 3w(1+w)$, whereas our model always has $w^\prime < 3w(1+w)$.  In this respect, it more
closely resembles the barotropic models discussed in refs.~\cite{wScherrer,LS}.
This
result arises from the fact that we have a two-component dark energy model.
In terms of
the evolution of the equation
of state, our model resembles the barotropic ``wet dark fluid" model proposed in \cite{water1,water2}, with the important difference that the dark
radiation in our model appears only at late times.

\begin{figure}\epsfig{file=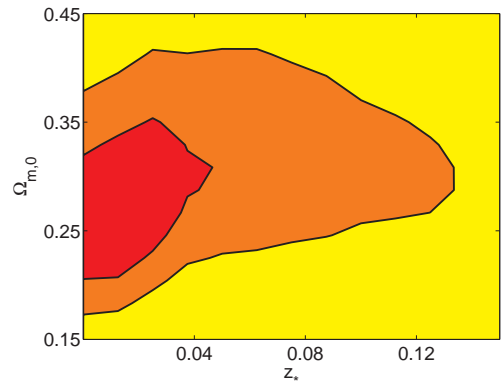,height=52mm}\caption{\label{zstar} Likelihood contour for the parameters $\Omega_{m0}$ and $z_*$, the redshift at which the Universe has exited the false vacuum and entered the true vacuum, releasing the energy of the cosmological constant into relativistic modes. The yellow (light) region is excluded at the 2$\sigma$ level, and the orange (darker) region is excluded at the 1$\sigma$ level. The red (darkest) region is not excluded at either confidence level.}
\end{figure}As a point related to our analysis, we consider the possibility that our Universe has exited the false vacuum in recent times, i.e.~all of the dark energy has recently been dumped into relativistic modes, which will eventually redshift away. This would be the case if the dark energy field went through a first-order phase transition of the type considered above, but into the true $V=0$ vacuum and not into another meta-stable vacuum. This scenario also arises in the ``accelerescence'' model considered in \cite{Chacko:2004ky}. Let $z_*$ be the redshift-time of this phase transition, when the vacuum energy is instantaneously (relative to cosmological timescales) converted into radiation. The Hubble parameter is therefore given by
\begin{equation}
\begin{split}
\label{hubbleexpansion}
H^2\(z\)= H_0^2\[\Omega_{m0}\(1+z\)^3+\Omega_{r0}\(1+z\)^4+\right.\\
\left.+\Omega_{\phi 0}\(1+z S\(z_*-z\)\)^4\]~,
\end{split}
\end{equation}
where $S\(x\)$ is the Heaviside step function. Using the SNIa data, \fig{zstar} is a likelihood plot for the parameters $\Omega_{m,0}$ and $z_*$. We find that $z_*$ is tightly constrained by the data, with the maximum $z_*$ allowed being $\sim0.1$ at $2\sigma$. Thus, if the Universe has already exited the vacuum energy epoch, it did so very recently.\vspace{-3mm}\subsection{Particle physics}\vspace{-1mm}An interesting feature of our scenario is that the dark energy field can be coupled relatively strongly to Standard Model particles. This makes it possible, in principle, for this kind of dark energy to be detected in colliders.\footnote{The proposed dark energy field $\phi$ (dark radiation) has very small mass $\sim{\rm meV}$, and might be produced by thermal reactions in stars. The energy loss argument for globular-cluster stars or red giants sets some strict limits on its coupling to the Standard Model, similar to constraints on the axion decay constant \cite{raffelt}. E.g., an ${\cal O}(1)$ Yukawa coupling of a scalar field $\phi$ to quarks $q$ induces, in one-loop, an effective dimension-5 coupling to photons $\alpha\phi\left(F_{\mu\nu}\right)^2/4\pi m_q$; this coupling would cause globular clusters to lose energy (into $\phi$ modes) more quickly than is actually observed, unless suppressed by a quark mass scale $m_q>10^7\,{\rm GeV}$, thus disallowing such a coupling to any Standard Model fermions. On the other hand, if the dark energy field obeys a $Z_2$ symmetry $\phi\to-\phi$, or if $\phi$ is the glueball field of some additional $SU(N)$ gauge theory (with interpolating dimension-4 operator $G_{\mu\nu}G^{\mu\nu}$) coupled to the Standard Model via messengers $m_q$, the induced effective interaction with photons has higher dimension; dimension-6 interactions $\alpha\phi^2\left(F_{\mu\nu}\right)^2/4\pi m_q^2$ can already avoid the energy-loss constraints for helium-burning stars ($T_{\rm core}\sim10^8\,{\rm K}$) if $m_q\rlap{\lower2pt\hbox{$\sim$}}\raise2pt\hbox{$>$}\,20\,{\rm GeV}$, thereby allowing coupling of the dark energy field to weak-scale Standard Model particles. These astrophysical constraints do not significantly hinder detection of our proposed dark energy field $\phi$ at particle colliders, which provide energies $\gg T_{\rm core}$ and produce weak-scale particles abundantly.}

The simplest model we considered, comprised of a singlet scalar $\phi$, has some challenges, as a direct coupling between $\phi$ and the Higgs boson operator $H^\dagger H$ cannot be excluded. This would lead to significant radiative corrections to the $\phi$ potential parameters, making the model somewhat unnatural. However, if this fine tuning is ignored, the $\phi$--$H$ coupling would provide for direct production of $\phi$ particles at colliders.

Our alternative model uses a pure $SU(N)$ gauge theory sector ($N>2$) with strong coupling scale $\Lambda \sim \Delta$. This model requires no fine tuning and the fraction of energy in relativistic modes after the phase transition can in principle be calculated from simulations of the $SU(N)$ theory. Glueballs of this sector would be light excitations with mass of order $\Delta$; the phase transition temperature would be at least a few times the glueball mass. The glueballs could couple to Standard Model particles via higher dimension operators such as
\begin{equation}
G_{\mu \nu}^2 \, O_{\rm sm}~,
\end{equation}
where $G$ is the $SU(N)$ field strength and $O_{\rm sm}$ a (Lorentz scalar) Standard Model operator such as $H^\dagger H$, $\bar{q} q$, etc. 

\begin{figure}
\epsfig{file=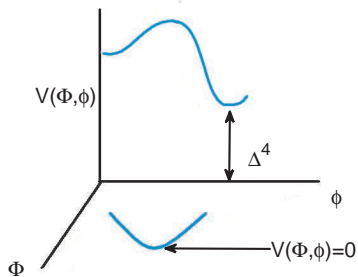,height=52mm}
\caption{\label{3dpotential} Potential energy surface for a gauge theory, where $\phi$ is an order parameter for confinement (Polyakov loop) and $\Phi$ a colored scalar field. For $N > 2$, $SU(N)$ models will have a first order confinement-deconfinement transition as the temperature is lowered. However, at zero temperature the deconfined phase ($\Phi = \phi = 0$) is not necessarily metastable.}
\end{figure}

If we wish to ensure that there exists a point in the configuration space where the vacuum energy vanishes exactly, one must add some extra degrees of freedom. For example, a colored scalar $\Phi$, whose potential $V \( \Phi \)$ has positive second derivative at $\Phi = 0$ and a global minimum at some non-zero value, would suffice (see \fig{3dpotential}). Note, this likely requires a non-renormalizable potential (i.e., with $\Phi^6$ term).

An exactly vanishing potential energy at some point in the configuration space is a generic feature of many theories with global supersymmetry \cite{Wittensusy} -- the vacuum energy is zero precisely at the supersymmetric points. While this fact does not explain away the cosmological constant term in the Einstein-Hilbert action, it may have something to do with the existence of an absolute minimum with small or vanishing energy density. In the supersymmetric framework, a presently non-zero and positive vacuum energy can be explained by the fact that the Universe is currently sitting at a meta-stable vacuum of the field theory, and its difference to a supersymmetric vacuum gives the present positive vacuum energy density $\Delta^4$. Examples of supersymmetric theories with such meta-stable vacua can readily be given either as simple Wess-Zumino models (e.g.~\cite{RayWZmodels}) or in terms of supersymmetric gauge theories which provide an ultraviolet framework for (O'Raifeartaigh-like) meta-stable supersymmetry breaking \cite{iss}. Furthermore, the dynamics for the Universe to initially be stuck in a meta-stable vacuum in the course of its cooling with only subsequent transition to the absolute supersymmetric minimum has been confirmed \cite{Universecooling} and the lifetime of such meta-stable vacua has been considered.

The phenomenologically plausible scale for electroweak supersymmetry breaking of $1\,{\rm TeV}\gg\Delta$ is much too large to directly account for the observed vacuum energy in the way just outlined. Nevertheless, supersymmetry might be invoked to provide for an absolute zero of the energy, at least in the sector containing the dark energy dynamics itself. By small modifications to the toy models described in the previous paragraph (changing parameters, or adding one or two new chiral superfields) it is furthermore possible to build dark energy sectors, containing just a small number of chiral superfields (possibly arising as effective fields \cite{iss}),  with \emph{two} slightly non-degenerate meta-stable vacua of energy $\sim\Delta$ along with a supersymmetric minimum, thereby giving a natural explanation for an absolute zero energy and also exhibiting the interesting dynamics of dark energy and recent dark radiation of our models (\ref{L}).

The dark energy sector is likely to feel electroweak supersymmetry breaking, at least through gravitational effects \cite{Chacko:2004ky}, and is therefore not expected to be perfectly supersymmetric. If this mediation happens only through gravitational interactions, the terms induced in the dark energy sector are naturally of the correct order of magnitude $(1\,{\rm TeV})^2/M_{\rm Pl}\sim10^{-3}\,{\rm eV}\sim\Delta$ to provide for energy differences in the dark sector of the size of the observed cosmological constant. Furthermore, a slight modification of the first model of \cite{Chacko:2004ky}, e.g.~addition of an $m\Phi^2$ term to the superpotential of the dark sector, generically yields three \emph{non}-degenerate meta-stable vacua of energy $\sim m\sim\Delta$, again yielding our scenario of dark radiation along with dark energy. If it is implemented in Nature, there could be dark radiation according to (\ref{hubble1}) as well as dark energy of order $\Delta$ present. And unlike in the model in \cite{Chacko:2004ky}, this scenario would cosmologically be detectable not solely in the very far future (billions of years from now), but could be confirmed, rejected or constrained already in the foreseeable future through comparison of the more precisely measured \emph{past} expansion rate of the Universe (for $0.1<z<3$) to our predictions for the equation of state $w(z)$ as Fig.~(\ref{wz}).

\section{Conclusions}
\label{Conclusions}

We have discussed a class of dark energy models which have interesting cosmological as well as collider signatures. In these models, a first-order phase transition at redshift $z\sim3$ releases energy in relativistic modes (dark radiation) leading to a characteristic time-dependence in the effective dark energy equation of state. We have shown that such models are consistent with SNIa data, and are relatively easy to construct as extensions to the Standard Model.

As an interesting and important side issue, we considered the possibility that the Universe might have recently (at redshift $z=z_*$) exited  the false vacuum phase and entered the true vacuum, converting all of the dark energy into relativistic modes. We show that the SNIa data places tight constraints on $z_*$, restricting it to $z_*\sim0.1$ or less at the $2\sigma$ confidence level. 
\acknowledgments
S.D.~acknowledges the hospitality of the Institute of Theoretical Science, University of Oregon. S.D.~and R.J.S.~were supported in part by the Department of Energy under No.~DE-FG05-85ER40226. S.D.H.H.~and D.R.~were supported by the Department of Energy under No.~DE-FG02-96ER40969.


\end{document}